\documentstyle[aps, prl, twocolumn]{revtex}
\newcommand{\be}{\begin{equation}}
\newcommand{\ee}{\end{equation}}
\newcommand{\bea}{\begin{eqnarray}}
\newcommand{\eea}{\end{eqnarray}}
\newcommand{\beas}{\begin{eqnarray*}}
\newcommand{\eeas}{\end{eqnarray*}}
\newcommand{\bi}{\begin{itemize}}
\newcommand{\ei}{\end{itemize}}
\newcommand{\bc}{\begin{center}}
\newcommand{\ec}{\end{center}}
\newcommand{\bfl}{\begin{flushleft}}
\newcommand{\efl}{\end{flushleft}}
\newcommand{\bfr}{\begin{flushright}}
\newcommand{\efr}{\end{flushright}}

\def\6{\partial}

\begin{document}

\preprint{ICEN xxxx/2001}

\title{Recovery Operator for Local Errors Generated by Gauge Multiplets} 
\author{Ion V. Vancea \thanks{Email: {\tt ivancea@ift.unesp.br,ion@gft.ucp.br}}}
\address{
$^1$ Instituto de F\'{\i}sica Te\'{o}rica, Universidade 
Estadual Paulista,
(IFT-UNESP)\\
Rua Pamplona 145, CEP 01405-900 S\~{a}o Paulo SP, Brasil\\
and\\
Centro Brasileiro de Pesquisas F\'{\i}sicas (CBPF)\\
Rua Dr. Xavier Sigaud 150, CEP 22290-180 Rio de Janeiro RJ, Brasil}
\maketitle

\begin{abstract}
The components of a quantum computer are quantum subsystems which have
a complex internal structure. This structure is determined by short-range interactions
which are appropriately described in terms of local gauge fields of the first kind.
Any modification of their state would produce, in general,
a new type of internal error, called local error, in the quantum state of 
the computer. We suggest that the general treatement of the 
local errors produced by a gauge multiplet can be done in the framework of 
the Algebraic Quantum Field Theory. A recovery operator is constructed from 
the first principles.  
\end{abstract}


Quantum computers are operational providing the errors in stocking and
manipulating the information are known and can be controled. Classifying and
correcting the quantum errors is crucial for practical reasons as
well as for defining rigorously an Universal Quantum Computer. Most of the errors 
that alter the memory register or the output of the quantum
computation are due to the small scale structure of the computer subsystems and to their
sensitivity to the interaction with the environment and with each other. The 
decoherence of the quantum state  and the uncertainty in the
unitary evolution have been studied for spin $\frac{1}{2}$-systems 
\cite{ps,ams,wz,dd,bdsw,cs,crss,lmpz,dg1,as,kl} and higher spin systems 
\cite{ek1,hfc1,hfc2,emr1}. Other effects that can alter the information as 
the quantum chaos \cite{ssb,gs,vvf}, the self-interaction among various qubits
\cite{gs,jgb} and the changes in the continuous variables of the system
\cite{slb,sljjs,jpwz2,slslb} have been considered and the codes to correct
them have been written down. A general observation that can be drawn from the
above studies is that long-range (electromagnetic) and/or non-linear interactions
are responsible for the alteration of the quantum state of the processing unit. This
statement resumes the idea that the quantum computers are "small, sensitive and
easily perturbed". However, one may ask whether the short-range interactions  
affect in any way the precision of quantum computations. 
At Quantum Mechanics scale ( $10^{-15} $m), the  external errors and the ones 
provocated by the preparation 
of the initial state, the manipulation of the computer or the reading of the final state 
are discussed  by assuming that the "computer parts"  (atoms, molecules or nuclei) 
are stable systems. Nevertheless, these components are, at their turn, quantum objects 
with an internal structure equally determined by short-range interactions. Even if, in
most of the practical situations, their life-time is large enough, they can have their 
internal state changed (like in spontaneous emission or decayment) . The modification
of the internal state of any subsystem of a computer may, in general, alter the state of 
the whole system inducing a new type of quantum errors. The purpose of this letter is to 
put this question into a more formal frame and to discuss the possibility of constructing 
recovery operators for this type of errors. 

The parts of a quantum computer are complex subsystems hold together by internal 
interactions. If we focus to nuclei, which form subsystems of any quantum computer
considered up to now, the most important role is played by gauge fields which produce
local interactions and are nonlinear. In general, a nucleus 
can be described as a state of the gauge field. Consequently, one can define
the state of the computer as a (more complicate) state in the Hilbert space 
of the field theory. A change of the internal state of a nucleus as, for example, a spontaneous 
emission that would carry away spin quantum number, may induce a modification of the 
quantum state of the computer. This transition can occur under the action of an external
system, as has been considered up to now, or of gauge fields. We are going to focus our
attention on the later case.   In the space representation, the transition of a nucleus from an
internal state to another is a local phenomenon due to local (short-range) fields. Therefore,
it is natural to assume that: i) there are local errors produced by local field operators in a 
local quantum gauge field theory and ii) the state of the computer is described by a vector in
the Hilbert space of the field theory. The physics of the field theory is contained in the
observables which are constructed  from field operators. 

There are three basic principles that such of theory should obey :
{\em the principle of locality}, {\em the principle of causality} and {\em the principle of
gauge invariance}. They guarantee that the errors in the quantum computer are local 
concearning a single subsystem of the computer and that the quantum field theory we are 
dealing with is a known one. Formally, the principle of locality states that to any open 
and finite extended region of space-time $\Sigma$ one can associate a *-algebra of observables
${\cal{O}}(\Sigma)$ and a *-algebra of fields ${\cal{F}}(\Sigma)$. 
(The involution * is necessary in order to define the conjugates of fields. In what follows we will 
denote it by $^{\dagger}$ as is usual.) One can define global algebras over any finite 
volume of the computer or any of its subsystems by considering the norm closed unions 
of the open neighbourhoods covering that region
\be
{\cal{O}} = \cup {\cal{O}}(\Sigma)~~~,~~~{\cal{F}} = \cup {\cal{F}}(\Sigma).
\label{global}
\ee
The principle of causality states that for any two space-like separated regions of space-time
$\Sigma_1$ and $\Sigma_2$ the corresponding algebras of observables commute with each other, i.e.
$\left[ {\cal{O}}(\Sigma_1 ), {\cal{O}}(\Sigma_2 ) \right] = 0$ (the observables must
be bosonos, but the fields may be fermions.)  The principle of gauge invariance tells us that
there is a representation of a gauge group $\cal{G}$ (denoted for simplicity with the same 
letter) that acts on the algebra of fields ${\cal{F}}(\Sigma )$. The algebra 
${\cal{O}}(\Sigma)$ is gauge invariant. Since we assume that the observables are constructed 
out of fields, ${\cal{O}}(\Sigma)$ is the gauge invariant part of the algebra ${\cal{F}}(\Sigma)$. 
If we want to describe the effects of short-range interactions we can limit ourselves to the 
gauge groups $SU(2)$ or $SU(3)$. However, we are going to  keep the discussion more general 
by working with arbitrary compact gauge groups. In what follows $\cal{G}$ denote
a compact gauge group that defines a gauge symmetry of the first kind, 
i. e. which excludes the long range interactions. Under these assumptions,
 the setting of the problem
is the Algebraic Quantum Field Theory  \cite{hk}. As was shown in 
\cite{dhr1,dhr2,dr1,dr2} the algebra
$\cal{F}$ is generated by a gauge multiplet $\{ \Psi_i \}$, $i=1,2,\ldots,n$ and $\cal{O}$. 
From the principle of locality and the principle of causality it follows that the multiplet 
should generate a 
Cuntz algebra $O_d$ \cite{dhr1,dhr2}, i.e. it satisfies the following relations \cite{dhr1,dhr2}
\bea
\sum_i\Psi_i {\Psi}^{\dagger}_i &=& 1\nonumber\\
{\Psi}^{\dagger}_i\Psi_j &=&\delta_{ij}.
\label{cuntzalgebra}  
\eea
There is a canonical endomorphism of the algebra $\cal{O}$ which defines the density matrix for 
any field $A \in \cal{F}$
\be
\rho (A) = \sum_i \Psi_i A \Psi^{\dagger}_i
\label{densitymatrix}
\ee
and which satisfies the following relation for any $\Psi$ and any $A$ from $\cal F$
\be
\Psi A = \rho (A) \Psi .
\label{propdensity}
\ee
It has been shown in \cite{dhr1,dhr2,dr3} that the mathematical structure of $\cal F $ is
that of a cross-product algebra of $O_d$ by the action of $\cal G$.

Now let us assume that the computer is prepared initially in the pure state $\left| \phi_I \right>$ 
which is a vector in the Hilbert space of the field theory and is determined by the states 
of the computers
subsystems (e.g, nuclei). The local errors which we are considering here represent a 
new state obtained 
by acting on the initial state with products of the field operators, i. e. elements of the algebra 
$\cal{F}$. For simplicity, we will consider in what follows only the action of the generators 
$\{ \Psi_i \}$ of $O_d$. Thus, the system evolves to an error state $\rho_F$ given by
\be
\rho_F = \sum_i \Psi_i \rho_I \Psi^{\dagger}_i ,
\label{finalstate}
\ee
where $\rho_I$ is the density matrix associated to the initial vector state.
We remark from (\ref{cuntzalgebra}) that the set of operators $\{ \Psi_i \}$ form a 
superoperator \cite{klbasic}. 
Due to this structure, it is appropriate to correct the error by using the  
recovery operator method, i.e. to construct a set of operators $\{ R_a \}$ that 
projects the wrong state $\rho_F$ back to the initial state \cite{klbasic}.  To this end,
note that one can associate a projector $P_i$ to each field operator $\Psi_i$ defined 
by the following relation
\be
P_i = \Psi_i \Psi^{\dagger}_i~~~,~~~P^2_i = P^{\dagger}_i = P_i,
\label{projector}
\ee
with the following action on any density matrix  
\be 
\rho ~ \mapsto ~ \rho_P = \sum_i P_i \rho P_i.
\label{actionproj}
\ee 
The relations (\ref{cuntzalgebra}) imply that $P_i$'s project on to orthogonal
directions in the Hilbert space 
\be
P_i P_j = \delta_{ij} P_j.
\label{orthoproj}
\ee 
The full set $\{ P_i \}$ projects the state $\rho_F$ to itself and map 
any other state $\rho\neq\rho_F$ to a different state.  Therefore, they are not 
suited to recover the initial state. However, since the states obtained by
projecting the final state through each $P_i$ are orthogonal on each other, one
could seek a linear combination of them that is the initial state of the computer.
Equivalently, one can look for a particular linear combinations of the
projectors that projects the final state on to the inital one.
This combination defines the recovery operator. 
The projected state is equal to  the initial state if its norm is equal to 
the norm of the initial state. Let us pursue these  ideas by requiring that the fidelity 
of the state obtained by acting with the recovery operator on $\rho_F$ 
be maximum in the direction of $\left| \phi_I \right>$ , i.e. equal to the norm 
of $\left| \phi_I \right>$. The fidelity of a state $\rho$ is defined as the square of 
the norm of the element matrix of it in the inital state
\be
F(\phi_I,\rho ) = \left< \phi_I \left| \rho \right| \phi_I \right>.
\label{fidelity}
\ee  
Consider the following linear combinations of projectors
\be
R_a = \sum_i \alpha_{ai}P_i,
\label{recoveryoperator}
\ee    
where $a=1,2,\ldots,M$ belongs to a discrete and finite set and $\alpha_{ai}$ are 
complex numbers.
$\{ R_a \}$ map $\rho_F$ to the state $\rho_R$ by the usual action of the 
operators on matrix densities
\be
\rho_F ~\mapsto ~\rho_R = \sum_a R_a \rho_F R^{\dagger}_a.
\label{recovact}
\ee  
The initial state is recovered by $R_a$'s if 
\be
F(\phi_I,\rho_R) = \left< \phi_I \left|\rho_I \right| \phi_I \right>.
\label{recovstate}
\ee
Assume for simplicity that the norm of the initial state is one. Then 
(\ref{recovstate}) is equivalent to the following relation 
\be
\sum_a\sum_i \left| \alpha_{ai} \right|^2 \left| \Psi_{iI} \right|^2
=1,
\label{constr1}
\ee
where $\Psi_{iI} = \left< \phi_I \left| \Psi_i \right| \phi_I \right>$.
The relation (\ref{constr1}) represents a constraint on the moduli of
the complex coefficients $\alpha_{ai}$ in terms of the known gauge multiplet
and the initial state. If a whole subset $C$ of the Hilbert space is
recovered by the operators $R_a$'s, then (\ref{constr1}) should hold for
any $\left| \phi_I \right>$ from $C$. Suppose that $C$ form a linear subspace 
of finite dimension $k$ of the Hilbert space of the field theory. By picking 
up a basis $\{ \left| e_A \right> \}$ of it, one can determine the
constants $\alpha_{ai}$'s, and consequently the operators $R_a$'s, up to some
phase factors. This can be achieved by  requiring that any of the elements of 
the basis be recovered by $R_a$'s and solving the corresponding linear system
\be
\sum_a\sum_i \left| \alpha_{ai} \right|^2 \left| \Psi_{iA} \right|^2
=1~~~,~~~A=1,2,\ldots,k. 
\label{constr2}
\ee
The number $M$ of the operators $R_a$ depends on the dimension $k$ of $C$ and
one should seek for its minimum value $M=k$ for which there are sufficient
equations in (\ref{constr2}) to determine the moduli of the coefficients 
$\alpha_{ai}$. 

The relations (\ref{constr2}) defines a linear-antilinear form on the vectors of the
subspace $C$. We would like to know what the transformation of the 
basis $\{ \left| \epsilon_A \right> \}$  that preserve (\ref{constr2}) are. Consider an
arbitrary linear transformation 
\be
\left | f_B \right> = \sum_C \theta_{BC} \left| e_A \right>,
\label{basistr}
\ee
where $B,C = 1,2,\ldots,M$ and $\theta_{BC}$ are the complex elements of the transformation
matrix between the two basis. If the relations (\ref{constr2}) are to hold for any of the elements
of the new basis on can find by an elementary algebra the following relation
\be   
\sum_{C} \left| \theta_{BC} \right|^2 + 
\sum_{C \neq D} \theta_{BC}^* \theta_{BD} \left| \Psi_{iCD} \right| = 1,~
\forall B=1,2,\ldots M,
\label{transfmatr}
\ee
where $\left| \Psi_{iCD} \right|  = \left< f_C \left| \Psi_i \right| f_D \right>$. The equation
(\ref{transfmatr}) defines the relations among the elements $\theta_{BC}$ of the transformation
matrix and imposes constraints on the sign of the biproducts $\theta_{BC}^*\theta_{CD}$.
Indeed, since  $\left| \Psi_{iCD} \right|  $ represents the probability of transition from the
state $\left| f_D \right>$ to the state $\left| f_C \right>$ under the action of the field operator
$\Psi_i$, it is a positive real number or zero. Consequently, if 
$\sum_{C} \left| \theta_{BC} \right|^2 =1 $ for any $B= 1,2,\ldots,M$, 
then the signs of the products $\theta_{BC}^*\theta_{CD}$ must alternate 
or $\left| \Psi_{iCD} \right|  $  must be all zero. Similar anaysis can be done if the first term in
(\ref{transfmatr}) is smaller or greater than one. The transformations that satisfy 
(\ref{transfmatr}) form a group. The recovery operator is invariant under these 
transformations, i. e. the coefficients $\alpha_{ai}$ do not depend on the vectors of
the code space $C$, but only on the gauge multiplet.

There is one more constraint that can be imposed naturally on the recovery
operator and it comes from the gauge structure of the theory. Firstly, note that 
the fidelity is gauge invariant 
\be
F(\phi_I',\rho_R') = F (\phi_I,\rho_R),
\label{gaugeinvfid}
\ee
where the transformations of the states and fields is given by the following relations
\bea
\left| \psi \right> & \mapsto & \left| \psi ' \right> = g \left| \phi_I \right>
\label{gaugetrstate}\\
\Psi_i & \mapsto & \Psi_i' = g\Psi_i g^{-1}
\label{gaugetroper}
\eea
for any $g \in \cal{G}$. Since the norm of the initial state is gauge invariant, the
equations (\ref{constr1}) and (\ref{constr2}) are gauge invariant, too, as expected.
Consider next the following isometry of the algebra $\cal{F}$ \cite{dr3}
\be
S = \frac{1}{\sqrt{d!}}\sum_{q\in P(d)} {\mbox{sign}} (q) \Psi_{q(1)} 
\Psi_{q(2)}\ldots\Psi_{q(d)},
\label{isometry}
\ee
where $P(d)$ is the permutation group of $d$ elements. $S$ is a gauge invariant object
of $\cal F$ \cite{dr2,dr3} and it acts on the initial state as
\be
\rho_I \mapsto \rho_S = S\rho_I S^{\dagger}.
\label{actionS}
\ee
Now consider the following transformations: act firstly with a gauge
transformation (\ref{gaugetrstate}) and (\ref{gaugetroper}) and then apply the gauge
invariant isometry (\ref{isometry}). Since $S$ is an gauge invariant operator, we 
require that the recovery operator $R_a'$ obtained from (\ref{gaugetroper})
maximizes the fidelity on the gauge transformed initial state $\left| \phi_I ' \right>$
\be
F(\phi_I',\rho_{RS}') = \left< \phi_I' \left|\rho_S' \right| \phi_I' \right>.
\label{recovgaugestate}
\ee
It is an elementary exercise to show that (\ref{recovgaugestate}) is equivalent to the same
equation for the untrasformed objects. After a simple algebra we obtain the following relation
from (\ref{recovgaugestate})
\bea
\sum_{a=1}^{M}& &\sum_{q,r \in P(d)}\frac{1}{d!} {\mbox{sign}}(q){\mbox{sign}}(r)
\alpha_{aq(1)}\alpha^{*}_{ar(1)}\times \nonumber\\
&&\left< \Psi_{q(1)}\cdots\Psi_{q(d)} \right>_I
\left< \Psi^{\dagger}_{r(d)}\cdots\Psi^{\dagger}_{r(1)} \right>_I
=1,
\label{constr3}
\eea
where $\left<\cdots\right>_I$ is a shorthand notation for 
$\left< \phi_I\left| \cdots \right| \phi_I \right>$. We interpret 
(\ref{constr3}) as a constraint on the complex coefficients $\alpha_{aI}$.

The equations (\ref{constr1}), (\ref{constr2}) and (\ref{constr3}) represent the
main result of the present work. They are based on the ansatz  
(\ref{recoveryoperator}). While (\ref{constr1}) and (\ref{constr2}) provide the
linear equations for determining the coefficents in the recovery operator 
${\cal{R}} = \{ R_a \}$ for $a=1,2,\ldots,M=k$ up to some phase factors,
the relation (\ref{constr3}) is obtained from the gauge symmetry of the
field operators. It states that under an arbitrary gauge transformation 
and a gauge invariant transformation the recovery operator  remains so. 
Since the phases of the complex coefficients $\alpha_{ai}$ are not 
constrained by any relation, we may fix them to zero to get real coefficients.
The dimension of the code space $C$ fixes the number of operators $R_a$. 
The equations (\ref{constr1}), (\ref{constr2}) and 
(\ref{constr3}) were determined from general principles of the Algebraic Quantum
Field Theory and from the requirement that the fidelity is maximized for any state
from the code space.  The relation (\ref{transfmatr}) defines the group of 
transformations that preserve the recovery operator $\cal R$ in the sense that it is
determined only by the gauge multiplet and not by the initial states of the system.
The interpretation of $\cal{R}$ is that of the recovery operator for the internal errors
produced by a local gauge multiplet. They alter the state of the computer either 
by changing the states of quantum subsystems or by changing locally the general state
of memory register. The physics behind the present description is that of modifications in
the internal state of the subsystems of the computer induced by short-distance interactions. 
Mathematically, we have been working in the frame of the Algebraic Quantum Field Theory. 

The discussion about the errors induced by the algebra of observables 
$\cal{O}$ stays open. On general grounds, we would expect that they produce similar 
errors as
the local fields. However, the algebra of observables does not have, in general, the 
structure of the Cuntz algebra. Another interesting class of internal error is that produced
by non-linearities in the gauge fields that are responsible for the internal structure of 
the quantum components of the computer. We hope to clarify  these issues somewhere else 
\cite{ivv}.  

I would like to thank J. A. B. de Oliveira for general discussions and to 
J. A. Hel\"{a}yel-Neto for comments and for his warm hospitality at 
GFT-UCP and DCP-CBPF during the 
period of ellaboration of this work. I also acknowledge a FAPESP postdoc 
fellowship.

\end{document}